\documentclass[preprint2]{aastex}

\usepackage{graphicx}
\usepackage{graphics}
\usepackage{amsmath}
\usepackage{txfonts}
\usepackage{epsfig}
\usepackage{natbib}
\usepackage{times}
\usepackage{amssymb}
\usepackage{color}
\usepackage{multicol}

\newcommand\igrone{\object{IGR~J14043$-$6148}}
\newcommand\igrtwo{\object{IGR~J16358$-$4726}}
\newcommand\igrthree{\object{IGR~J16393$-$4643}}
\newcommand\igrfour{\object{IGR~J17091$-$3624}}
\newcommand\igrfive{\object{IGR~J17597$-$2201}}
\newcommand\ergcms{erg\,cm$^{-2}$\,s$^{-1}$}

\newcommand\ergs{erg\,s$^{-1}$}
\newcommand\countss{count\,s$^{-1}$}
\newcommand\cmsq{cm$^{-2}$}
\newcommand\integ{{\it{INTEGRAL}}}

\newcommand\rxte{{\it{RXTE}}}
\newcommand\swift{{\it{Swift}}}
\newcommand\asca{{\it{ASCA}}}
\newcommand\xmm{{\it{XMM-Newton}}}
\newcommand\chan{{\it{Chandra}}}
\newcommand\nh{$N_\mathrm{H}$}

\shorttitle{\emph{Chandra}-HRC observations of five \emph{INTEGRAL} sources}

\shortauthors{Bodaghee et al.}

\begin{document}

\title{\emph{Chandra} observations of five \emph{INTEGRAL} sources: \\ new X-ray positions for \object{IGR~J16393$-$4643} and \object{IGR~J17091$-$3624}}

\author{A. Bodaghee}
\affil{Space Sciences Laboratory, 7 Gauss Way, University of California, Berkeley, CA 94720, USA}
\email{bodaghee@ssl.berkeley.edu}

\and

\author{F. Rahoui}
\affil{Harvard University, Astronomy Department, 60 Garden Street, Cambridge, MA 02138, USA}

\and

\author{J. A. Tomsick}
\affil{Space Sciences Laboratory, 7 Gauss Way, University of California, Berkeley, CA 94720, USA}

\and

\author{J. Rodriguez}
\affil{Laboratoire AIM, CEA/IRFU - Universit\'e Paris Diderot - CNRS/INSU, \\ CEA DSM/IRFU/SAp, Centre de Saclay, F-91191 Gif-sur-Yvette, France}

\begin{abstract}
The \chan\ High Resolution Camera observed the fields of five hard X-ray sources in order to help us obtain X-ray coordinates with sub-arcsecond precision. These observations provide the most accurate X-ray positions known for \object{IGR~J16393$-$4643} and for \object{IGR~J17091$-$3624}. The obscured X-ray pulsar \object{IGR~J16393$-$4643} lies at R.A. (J2000) $=$ $16^{\mathrm{h}}$ $39^{\mathrm{m}}$ $05\overset{\mathrm{s}}{.}47$, and Dec. $=$ $-46^{\circ}$ $42^{\prime}$ $13\overset{\prime\prime}{.}0$ (error radius of $0\overset{\prime\prime}{.}6$ at 90\% confidence). This position is incompatible with the previously-proposed counterpart \object{2MASS~J16390535$-$4642137}, and it points instead to a new counterpart candidate that is possibly blended with the 2MASS star. The black hole candidate \object{IGR~J17091$-$3624} was observed during its 2011 outburst providing coordinates of R.A. $=$ $17^{\mathrm{h}}$ $09^{\mathrm{m}}$ $07\overset{\mathrm{s}}{.}59$, and Dec. $=$ $-36^{\circ}$ $24^{\prime}$ $25\overset{\prime\prime}{.}4$. This position is compatible with those of the proposed optical/IR and radio counterparts, solidifying the source's status as a microquasar. Three targets, \object{IGR~J14043$-$6148}, \object{IGR~J16358$-$4726}, and \object{IGR~J17597$-$2201}, were not detected. We obtained 3$\sigma$ upper limits of, respectively, 1.7, 1.8, and $1.5\times10^{-12}$\,\ergcms\ on their 2--10\,keV fluxes.

\end{abstract}


%%__________________________________________________________________
\section{Introduction}

Surveys by \integ\ have enabled the discovery of hundreds of new high-energy sources \citep[e.g.,][]{bir10,kri10}. While the soft $\gamma$-ray imager has proven adept at finding new sources dubbed \integ\ Gamma-Ray sources or IGRs\footnote{a comprehensive list can be found at \texttt{http://irfu.cea.fr/Sap/IGR-Sources}}, the position error radii are on the order of a few arcminutes. These are clearly too large to permit the identification of a single counterpart in the optical and infrared (IR) bands. Establishing the nature of the optical/IR counterpart is a key step in helping to categorize a gamma-ray source into one of the many groups of high-energy emitters.

%__________________________________________________________________Journal
%
\begin{table*}[!t] 
\caption{Journal of \chan\ observations of the five targets in this study.}
\vspace{2mm}
\begin{tabular}{ l c c c c }
\hline
\hline
Target	      				& Obs. ID 		& Start Time (UTC)		& Start Time (MJD)		& Exposure Time (s)		\\	
\hline
\object{IGR~J14043$-$6148}	& 12502		& 2010-12-06T02:42:53	& 55536.11311			& 1128.95 	\\	
\object{IGR~J16358$-$4726}	& 12503		& 2011-02-05T05:41:21	& 55597.23705			& 1142.88		\\
\object{IGR~J16393$-$4643}	& 12504		& 2011-03-05T03:12:11	& 55625.13346			& 1125.20		\\
\object{IGR~J17091$-$3624}	& 12505		& 2011-03-06T00:38:35	& 55626.02679			& 1124.89		\\
\object{IGR~J17597$-$2201}	& 12506		& 2011-02-20T11:33:00	& 55612.48125			& 1109.14		\\					
\hline
\end{tabular}
\label{tab_log}
\end{table*}

Therefore, the classification of these sources depends on observations with X-ray focusing telescopes which provide position accuracies of a few arcseconds \citep[e.g.,][]{rod10}. In some cases, these error radii encompass multiple potential counterparts, especially for X-ray sources located in crowded stellar fields such as the Galactic Center and Plane. In such cases, only a sub-arcsecond X-ray position obtained with \chan\ will help localize the correct counterpart enabling follow-up spectral studies to be performed in the optical/IR, which will eventually help lead to a source classification.

Here, we present the results from \chan\ snapshot observations of five IGRs located along the Galactic Plane: \igrone, \igrtwo, \igrthree, \igrfour, and \linebreak \igrfive. These objects were previously observed with \rxte, \swift, or \xmm, and so some of their spectral and timing behavior is known. They share a common trait in that the identity of the optical/IR counterpart was not firmly established when the \chan\ observations were proposed. These \chan\ observations are intended to help refine the positions to the X-ray sources. The observational history of our selected targets is summarized in the following paragraphs. Data and analysis methods are presented in Section\,\ref{sec_obs}. Results for detected sources are discussed in Section\,\ref{sec_det}, while the implications for undetected sources are raised in Section\,\ref{sec_undet}.

\subsection{IGR~J14043$-$6148}

This source was first listed in the 4$^{\mathrm{th}}$ ISGRI Catalog \citep{bir10}. Within the 4$\overset{\prime}{.}$5 error radius from ISGRI, lies an X-ray source from the 2$^{\mathrm{nd}}$ \xmm\ Serendipitous Source Catalog \citep{wat09}: \object{2XMM~J140417.3$-$614911}, which is only $0\overset{\prime}{.}4$ away from the centroid of the ISGRI position. Its position uncertainty of $2\overset{\prime\prime}{.}5$ is too large to permit the identification a single infrared counterpart in the 2MASS Catalog \citep{cut03,skr06}. 

A \swift-XRT observation of the field was performed on 2010 June 2 (i.e., around 6 months before our \chan\ pointing) revealing an X-ray source at R.A. (J2000) $=$ $14^{\mathrm{h}}$ $04^{\mathrm{m}}$ $29\overset{\mathrm{s}}{.}63$, and Dec. $=$ $-61^{\circ}$ $47^{\prime}$ $19\overset{\prime\prime}{.}7$ with a 90\%-confidence error circle of $4\overset{\prime\prime}{.}5$ \citep{lan11}. In other words, the XRT source is compatible with the ISGRI position, but it is $2\overset{\prime}{.}4$ away from (and incompatible with) the 2XMM source. Within the XRT error circle lies \object{G311.45-0.13} which has been proposed to be either a supernova remnant (SNR) or a background active galactic nucleus (AGN).

%__________________________________________________________________Chandra Image
\begin{figure*}[!t] 
\centering
\includegraphics[width=\textwidth,angle=0]{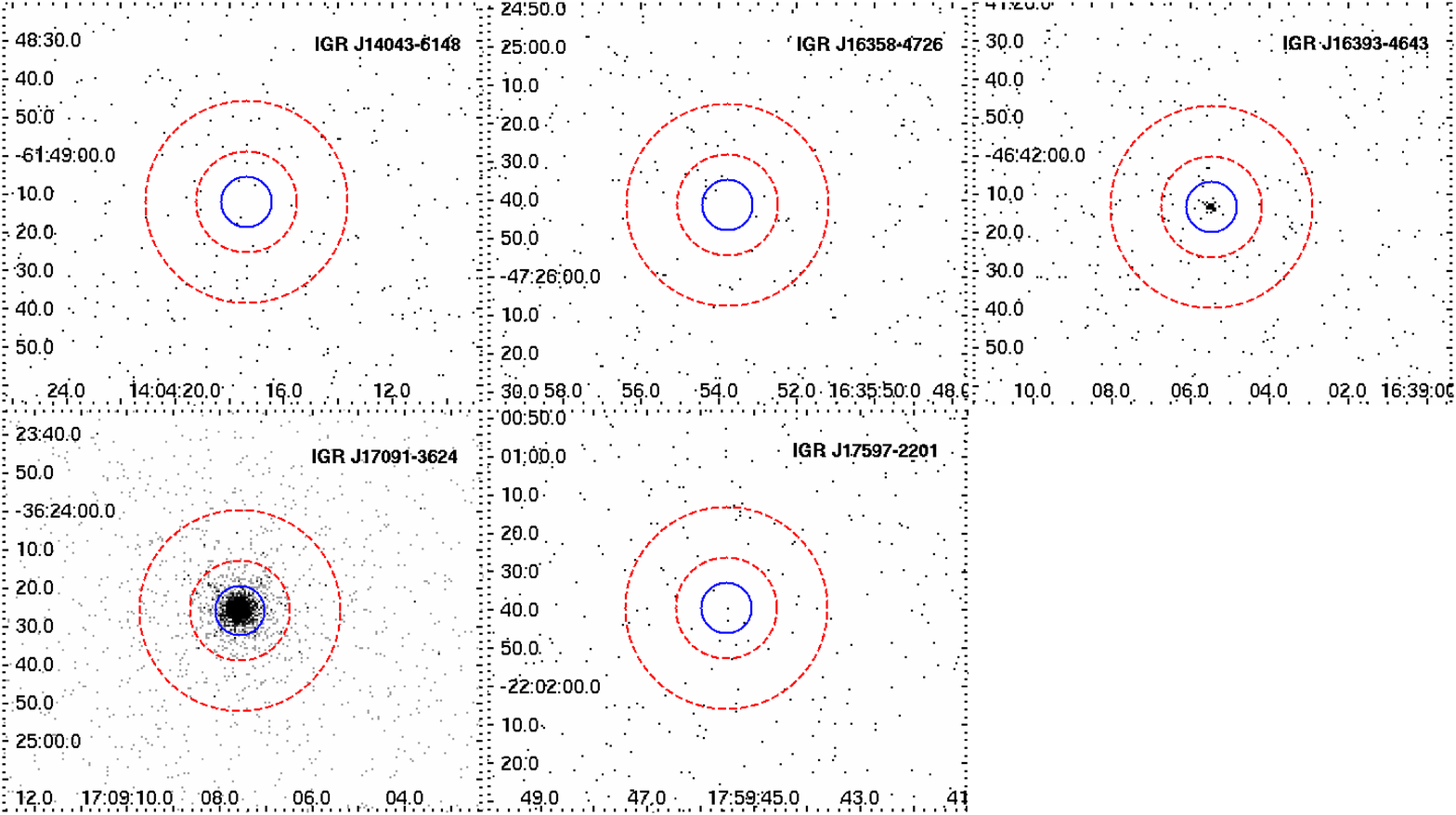}
\caption{\chan-HRC images (0.3--10\,keV) of the fields of the five targets in our study. Coordinates are given as equatorial R.A. and Dec. (J2000) where North is up and East is left. The circle (solid line) and annulus (dashed line) represent the source and background extraction regions, respectively. Pixels have been grouped in blocks of 4 and the scaling is logarithmic.}
\label{img_chan}
\end{figure*}

\subsection{IGR~J16358$-$4726}

Discovered early in the \integ\ mission by \citet{rev03}, \object{IGR~J16358$-$4726} was serendipitously observed by \chan\ during an observation of the field of \object{SGR~1627$-$41} \citep{kou03} and associated with \object{2MASS~J16355369$-$4725398}. \citet{pat04} discovered a coherent modulation period lasting nearly 6000\,s, which could either be the spin period of a neutron star in an absorbed ($N_{\mathrm{H}} \sim 3\times10^{23}$\,cm$^{-2}$) HMXB system, or the orbital period of a low-mass X-ray binary (LMXB). The latter was ruled out once \citet{pat07} measured an increasing frequency for the period which points to an accretion-induced spin up of a highly-magnetized neutron star ($B \sim 10^{13}$--$10^{15}$\,G), i.e., a magnetar. Infrared observations of the field by \citet{cha08} and \citet{rah08} suggest a faint, absorbed, and possibly blended counterpart with a spectral energy distribution (SED) typical of a supergiant B[e] star. This seemed to confirm the HMXB nature of \object{IGR~J16358$-$4726}. However, based on the presence of CO absorption lines in the $K_s$-band spectra of the 2MASS source, \citet{nes10} suggest that the neutron star is not paired with an OB supergiant but with a KM giant, making the system a symbiotic X-ray binary.

\subsection{IGR~J16393$-$4643}

Originally discovered by \asca\ \citep{sug01}, \object{IGR~J16393$-$4643} was re-discovered in high energies by \integ\ \citep{bir04}. The \asca\ and \integ\ positions, which fall within the error box of an unidentified \emph{EGRET} source, are consistent with the positions of sources from the radio and IR bands, leading to its initial classification as a microquasar candidate \citep{com04,mal04}. An \xmm\ observation refined the position to a few arcseconds, thereby ruling out the previously proposed identifications and pointing towards \object{2MASS~J16390535$-$4642137} as the counterpart \citep{bod06}. The optical/IR SED of \object{2MASS~J16390535$-$4642137} suggests a star of spectral class BIV-V \citep{cha08}. However, analysis of the $K_s$-band spectrum of the same object hint at a late-type KM star in a symbiotic binary system \citep{nes10}, so there is disagreement about how to interpret the infrared SED, and there are lingering doubts about whether the 2MASS source is really the counterpart: the \xmm\ error circle includes 3 other candidates. The orbital period of the X-ray source was recently determined to be 4.2\,d \citep{cor10} which is at odds with a symbiotic system \citep[see also][]{tho06}. This short orbital period paired with a slowly-rotating neutron star \citep[spin period $\sim$900\,s: ][]{bod06} are typical of wind-accreting HMXBs \citep{cor86,bod07}.

\subsection{IGR~J17091$-$3624}

An \integ\ Galactic Center Deep Exposure uncovered \object{IGR~J17091$-$3624} \citep{kuu03}. The refined X-ray position obtained by \swift\ \citep{ken07} ruled out previously-proposed radio and optical counterparts, leaving two blended candidate NIR counterparts in the error circle \citep{cha08}. Re-analysis of archival VLA data from this field revealed a new compact radio counterpart compatible with the \swift\ position \citep{cap09}.

In early 2011, \citet{kri11} announced that \swift-BAT detected an outburst from \object{IGR~J17091$-$3624}. Optical/IR observations of the field \citep{tor11} performed before and during the outburst showed variability from the location of ``Candidate 2'' (C2) of \citet{cha08}. \citet{tor11} concluded from PSF-fitting that C2 is actually composed of two stars, the brighter one of which is the likely counterpart: it has magnitudes $I=$18.35$\pm$0.03 and $K_{s}=$16.98$\pm$0.04, and a position of R.A. (J2000) $=$ $17^{\mathrm{h}}$ $09^{\mathrm{m}}$ $07\overset{\mathrm{s}}{.}62$ and Dec. $=$ $-36^{\circ}$ $24^{\prime}$ $25\overset{\prime\prime}{.}35$ (error of $\sim 0\overset{\prime\prime}{.}2$ at 90\% confidence).

During this latest outburst, radio emission was detected again further cementing the source's status as a microquasar consisting of an accreting black hole candidate captured in the low-hard state \citep{cor11,rod11}. Observations with \rxte\ unveiled low-frequency quasi-periodic oscillations \citep[QPOs: ][]{rod11} whose evolving timing signatures \citep{sha11} were indicative of a transition to the high-soft state \citep{del11}. Millihertz and high-frequency QPOs were discovered \citep{alt11a,alt12} in addition to heartbeat oscillations \citep{alt11b}, and seven of the twelve variability classes seen in \object{GRS~1915$+$105} \citep{alt11c, alt11d,pah11}. However, \igrfour\ is less radio bright than is \object{GRS~1915$+$105} \citep{rod11}, and it loops through the hardness-intensity diagram in the opposite direction \citep{alt11d}. Assuming it is emitting at the Eddington limit during its outburst \citep[in accordance with models that suggest the broad range of variability is exclusively due to disk instabilities at high luminosities:][and references therein]{alt11d}, then \object{IGR~J17091$-$3624} either hosts the black hole with the lowest mass known \citep{alt11d}, or it is located far away in the Galaxy with distance estimates ranging from 17\,kpc \citep{rod11} to more than 20\,kpc \citep{alt11d}.

%__________________________________________________________________Journal
%
\begin{table*}[!t] 
\caption{Positions (J2000), detection significance, and background-subtracted HRC-I count rates (0.3--10\,keV) of the IGR sources in this program. The position uncertainty is $0\overset{\prime\prime}{.}6$ at 90\% confidence. Upper limits (3$\sigma$) on the count rate are given for sources which were not detected. }
\vspace{2mm}
\begin{tabular}{ l c c c c c }
\hline
\hline
Source Name	      			& \emph{Chandra} Counterpart			& R.A. 		& Dec.			& Count Rate	& Significance \\	
\hline
\object{IGR~J14043$-$6148}	& ---								& ---			& ---	& $\leq$0.008 & ---	\\	
\object{IGR~J16358$-$4726}	& ---								& ---			& ---	& $\leq$0.005 & ---	\\
\object{IGR~J16393$-$4643}	& \object{CXOU~J163905.5$-$464213}	& $16^{\mathrm{h}}$ $39^{\mathrm{m}}$ $05\overset{\mathrm{s}}{.}47$		& $-46^{\circ}$ $42^{\prime}$ $13\overset{\prime\prime}{.}0$	& 0.033	& 12 \\
\object{IGR~J17091$-$3624}	& \object{CXOU~J170907.6$-$362425}	& $17^{\mathrm{h}}$ $09^{\mathrm{m}}$ $07\overset{\mathrm{s}}{.}59$		& $-36^{\circ}$ $24^{\prime}$ $25\overset{\prime\prime}{.}4$	& 37.050	& 776  \\
\object{IGR~J17597$-$2201}	& ---								& ---			& ---	& $\leq$0.009 & ---		\\			
\hline
\vspace{-5mm}

\end{tabular}
\label{tab_det}
\end{table*}

\subsection{IGR~J17597$-$2201}

\integ\ detected \object{IGR~J17597$-$2201} in outburst \citep{lut03}, and it was quickly associated with the transient X-ray source \object{XTE~J1759$-$220} \citep{mar03}. In the optical/NIR band \citep{cha08}, the \xmm\ position error circle encompasses multiple stars, any one of which could be the counterpart to this dipping X-ray burster \citep[i.e., a neutron star binary in a high-inclination orbit:][]{bra07}. A \chan\ observation performed in 2007 \citep{rat10} gave source coordinates of R.A. (J2000) $=$ $17^{\mathrm{h}}$ $59^{\mathrm{m}}$ $45\overset{\mathrm{s}}{.}52$ and Dec. $=$ $-22^{\circ}$ $01^{\prime}$ $39\overset{\prime\prime}{.}2$ with an error radius of $0\overset{\prime\prime}{.}6$ (90\% confidence). \citet{rat10} found a single IR-band object within the \chan\ error circle, and it is coincident with ``Candidate 1'' from \citet{cha08}. This object is probably a low-mass star (based on the properties of the X-ray source) and is the likely counterpart to \object{IGR~J17597$-$2201}.

%%__________________________________________________________________
\section{Observations \& Analysis}
\label{sec_obs}

\subsection{\emph{Chandra}}

Our target list consists of five \integ\ sources for which \chan\ could improve upon the position accuracy given by other X-ray imaging telescopes. The fields of these sources were observed for $\sim$1\,ks each by the High Resolution Camera \citep[HRC: ][]{mur97} aboard \chan. These observations were scheduled between 2010 December and 2011 March (PI: Bodaghee). Table\,\ref{tab_log} provides the observation logs.

Data reduction relied on CIAO 4.3 and HEAsoft 6.11. Each events file was rebinned in blocks of 4 pixels with \texttt{dmcopy}, and the same was done for the reprojected background events file. The exposure-weighted background events were then subtracted using \texttt{dimgcalc}. An aspect histogram was computed with \texttt{asphist}, an instrument map (adopting the default value of 1\,keV for the mono-energy parameter) was made with \texttt{mkinstmap}, and an exposure map was generated with \texttt{mkexpmap}. The tool \texttt{dmimgthresh} allowed us to select only those pixels with exposure times that were at least 1.5\% of the maximum exposure in the map. Using \texttt{dimgcalc}, we divided the image by this exposure-selected map in order to obtain an image whose units are ph\,cm$^{-2}$\,s$^{-1}$\,pixel$^{-1}$.

We ran \texttt{wavdetect} on the full HRC energy range (0.3--10\,keV) to create an output list of detected sources (if any), and we performed visual checks of each image. As a consistency check, we verified that the coordinates from \texttt{wavdetect} were statistically compatible with those obtained by centroid-fitting the pixel distribution with \texttt{dmstat}. We adopt a position uncertainty of $0\overset{\prime\prime}{.}6$ at 90\% confidence which is equivalent to the \chan\ boresight uncertainty for an on-axis source. All positions are given in the 2000.0 epoch. For detected sources, we extracted source counts with \texttt{dmstat} from a 50-pixel radius (in the binned image) around the optimal position found with \texttt{wavdetect}. Background counts were extracted from an annulus of between 100 and 200-pixel radius with the same center. When the target was not detected, we used the same source and background geometries, but we centered them at the best known X-ray position in the literature, and ran \texttt{aprates} to determine an upper-limit on the source flux assuming that 99\% of the PSF is contained within the source region, and 1\% is in the background region. Images from HRC (0.3--10\,keV) are presented in Fig.\,\ref{img_chan}. The results from \texttt{wavdetect}, as well as the area-scaled and background-subtracted flux (or upper limits) from each source, are summarized in Table\,\ref{tab_det}.

Finally, it is important to note that in images created from standard processing, a jet-like feature protrudes from \object{IGR~J17091$-$3624} to a location $\sim$5--8$^{\prime\prime}$ away in the NE direction. Readout streaks are not expected for the HRC, but image artifacts that resemble hooks and jets have been noticed in certain images of bright point sources due to electronic ringing signatures \citep{jud00,mur00,kap08}. The recommended correction for this effect is to exclude events where the amplification scale factor ``\texttt{AMP\_SF} $=$ 3.'' We generated an events file that excludes ``\texttt{AMP\_SF} $=$ 3'' events, and another file that contains only these events (27228 out of 138461 total events). The NE extension can only be seen in images produced from events files that include ``\texttt{AMP\_SF} $=$ 3'' events, suggesting that electronic ringing is responsible for its appearance. The position and flux measurements that we quote for \object{IGR~J17091$-$3624} in Table\,\ref{tab_det}, as well as the image shown in Fig.\,\ref{img_chan}, are based on the corrected events list.

\subsection{\emph{Infrared Data}}

On 2011 July 19, we observed \object{IGR~J16393$-$4643} in the framework of a Norma Arm deep-field near-IR survey with the National Optical Astronomy Observatory (NOAO) Extremely Wide Field Infrared Imager (NEWFIRM), in the $J$, $H$, and $K_{s}$ broadband filters. The instrument, mounted on the CTIO/Blanco 4\,m telescope, has a 28$^{\prime}\times28^{\prime}$ field of view (FOV) and a $0\overset{\prime\prime}{.}392$ plate scale. In each filter, we obtained $10\times15$\,s dithered frames, allowing an accurate median sky calculation in this very crowded field. We reduced the data using the dedicated NEWFIRM package available with the \texttt{IRAF} suite, following the standard near-IR procedure which includes the correction for bad pixels and dark current, flat fielding, and median sky subtraction. 

We then performed accurate astrometry (rms less than $0\overset{\prime\prime}{.}1$) with \texttt{GAIA} included in the \texttt{starlink} suite, fitting an astrometric solution using the 2MASS stars located in the entire NEWFIRM field. The images were eventually flux-calibrated through relative photometry with respect to the 2MASS catalog. We derived the following zero-point magnitudes and atmospheric extinction coefficients: $Zp_{J}=22.64\pm0.08$, $Zp_{H}=22.69\pm0.10$, $Zp_{K_{s}}=21.83\pm0.15$; and $E_{J}=0.098\pm0.031$, $E_{H}=0.062\pm0.021$, and $E_{K_{s}}=0.089\pm0.023$, respectively\footnote{The conversion from the instrumental ($m_{\rm inst}$) to apparent ($m_{\rm app}$) magnitude is $m_{\rm app}=Zp+m_{\rm inst}-E\times AM$, where AM is the average airmass during the integration.}.

%__________________________________________________________________
\section{Detected sources}
\label{sec_det}

\subsection{IGR~J16393$-$4643}

%__________________________________________________________________16393 Image
\begin{figure*}[!t] 
\centering
\includegraphics[width=\textwidth,angle=0]{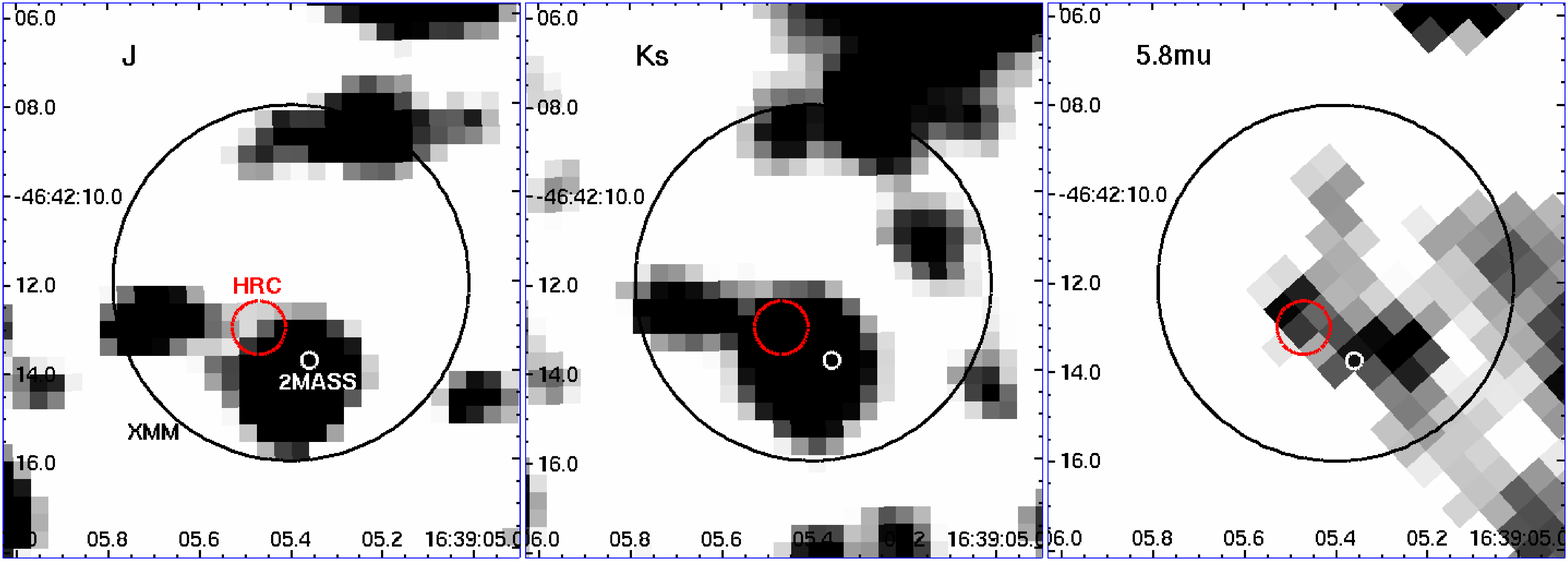}
\caption{Image of the field of IGR~J16393$-$4643 as captured by the NOAO/NEWFIRM telescope in the $J$ and $K_{s}$ bands (left and middle panels), and at 5.8\,$\mu$m (right panel) from \emph{Spitzer}-IRAC \citep{ben03}. Coordinates are given as equatorial R.A. and Dec. (J2000) where North is up and East is left. The error circle corresponding to the HRC position from this work, the \xmm\ position \citep{bod06}, and that of the previously-proposed infrared counterpart 2MASS~J16390535$-$4642137 \citep{cut03} are indicated. }
\label{img_16393}
\end{figure*}

We detected \object{IGR~J16393$-$4643} with HRC at a count rate of 0.033\,\countss\ (0.3--10\,keV), which corresponds to an absorbed flux of $1.5\times10^{-11}$\,\ergcms\ (2--10\,keV) assuming an absorbed power law with \nh $=2.5\times10^{23}$\,\cmsq\ and $\Gamma=0.8$ \citep{bod06}. This is around twice the average flux measured by ISGRI when translated to the same energy range \citep[$6.3\times10^{-12}$\,\ergcms,][]{bir10}.

Our \chan\ observation of \object{IGR~J16393$-$4643} provides the most precise position for the X-ray source: R.A. (J2000) $=$ $16^{\mathrm{h}}$ $39^{\mathrm{m}}$ $05\overset{\mathrm{s}}{.}47$ and Dec. $=$ $-46^{\circ}$ $42^{\prime}$ $13\overset{\prime\prime}{.}0$ with an error radius of $0\overset{\prime\prime}{.}6$ (90\% confidence). The HRC position is consistent with those obtained with \xmm\ and with \integ. However, this position is incompatible with the position of \object{2MASS~J16390535$-$4642137} which is located $1\overset{\prime\prime}{.}4$ away and which has an error radius of $0\overset{\prime\prime}{.}12$ (at 90\% confidence, see Fig.\,\ref{img_16393}). Thus, we conclude that {2MASS~J16390535$-$4642137}, whose optical/IR analysis led to diverging conclusions about its spectral class, is unlikely to be the counterpart to \object{IGR~J16393$-$4643}. None of the four candidate counterparts proposed in \citet{cha08} are consistent with the \chan\ position.

This suggests that the true counterpart to \object{IGR~J16393$-$4643} might be a distant (and reddened) star that is blended with the bright 2MASS star. We acquired deep near-IR images of the field of \object{IGR~J16393$-$4643} with the NOAO/NEWFIRM telescope, and combined these with archival \emph{Spitzer}-IRAC observations from the GLIMPSE campaign in the mid-IR \citep{ben03}. These images are presented in Fig.\,\ref{img_16393}. As the wavelength increases, the point spread function (PSF) for the bright 2MASS source is displaced in the direction of the \chan\ position. What appears as a single object in the $J$ band is revealed as two distinct peaks in the counts map of the 5.8\,$\mu$m band: one corresponds to the 2MASS object, and the other could be a deeply-reddened star situated on the wings of the PSF of the 2MASS star. At 8\,$\mu$m (not displayed), neither source candidate can be disentangled from the tail of a cloud-like region of diffuse emission.

If this candidate is real, and if it represents the true counterpart to \object{IGR~J16393$-$4643}, then it has been effectively isolated at 5.8\,$\mu$m. There are no objects listed in the GLIMPSE catalog \citep{ben03} consistent with the \chan\ position. The sensitivity limit of the GLIMPSE survey is 0.7\,mJy at 5.8\,$\mu$m \citep{chu09}. This limit would place an undetected supergiant O9 star ($T_{\mathrm{eff}} \sim 30,000$\,K, and $R \sim 20$\,$R_{\odot}$) with $A_{V} \sim 20$ at a minimum distance of 25\,kpc, which is probably too far to be plausible. On the other hand, for an undetected main-sequence B star ($T_{\mathrm{eff}} \sim 24,000$\,K, and $R \sim 10$\,$R_{\odot}$), the sensitivity limit implies a large, but reasonable distance ($\sim$12\,kpc away). We point out that \object{IGR~J16393$-$4643} is positionally coincident with the edge of an active, massive star-forming (H\,II) region situated at a distance of 12.0$\pm$0.3\,kpc \citep{rus03}. If the blended counterpart to \object{IGR~J16393$-$4643} is a main-sequence B star that originated from this H\,II region, then this would suggest a distance of $\sim$12\,kpc to the X-ray source, consistent with our estimate from the sensitivity limit.

\subsection{IGR~J17091$-$3624}

%__________________________________________________________________17091 Image
\begin{figure}[!t] 
\centering
\includegraphics[width=0.45\textwidth,angle=0]{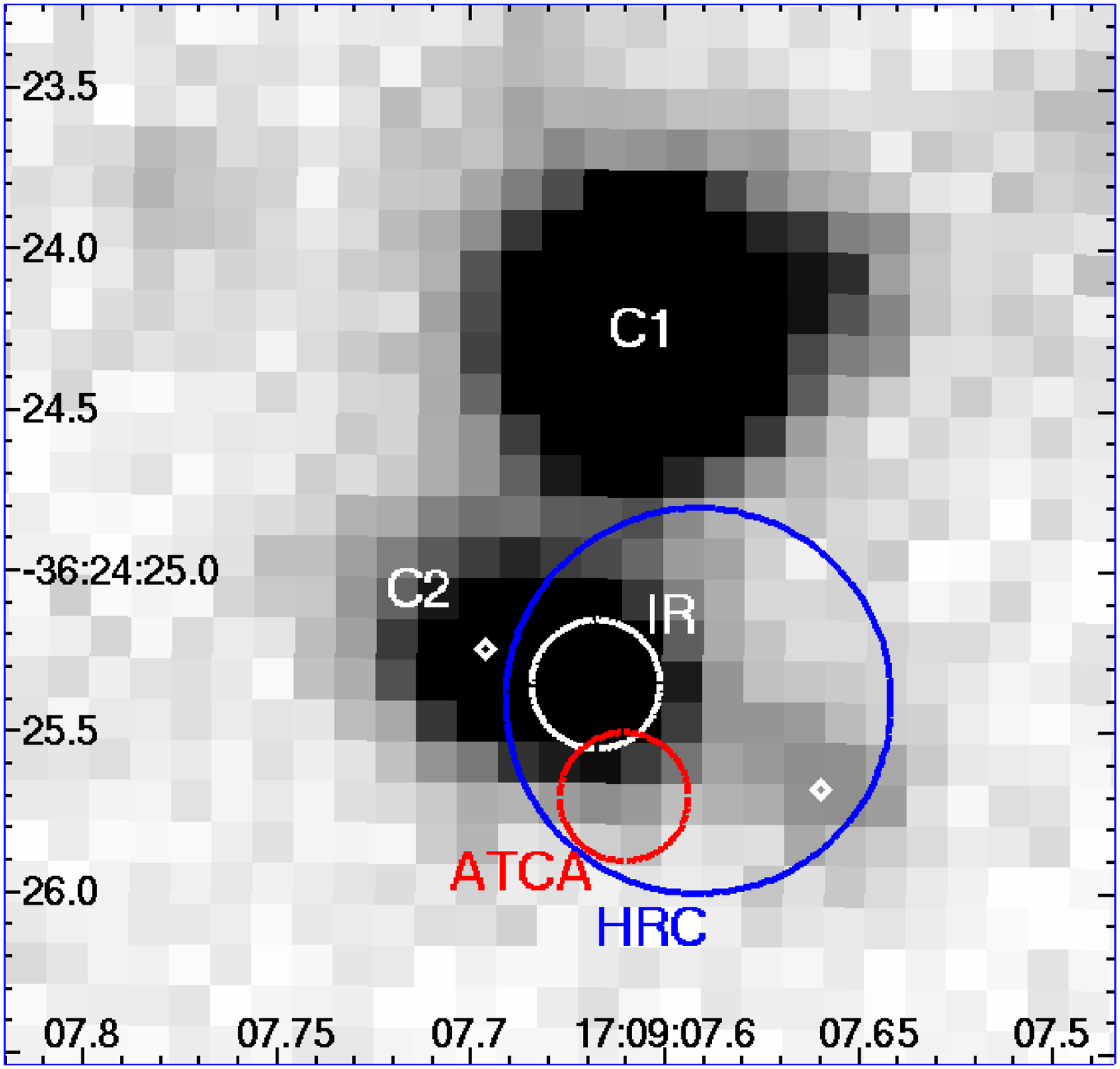}
\caption{The field of IGR~J17091$-$3624 in the $I$ band as captured by the IMACS imaging spectrograph on the 6.5-m Magellan Baade Telescope at the Las Campanas Observatory during 2011 Feb. 6 \citep{tor11}. Coordinates are given as equatorial R.A. and Dec. (J2000) where North is up and East is left. The error circles corresponding to the \chan-HRC position (this work), and that of the proposed infrared \citep{tor11} and radio \citep[ATCA: ][]{rod11} counterparts are indicated. The \swift-XRT error circle of  \citet{ken07} is larger than the image. Also indicated are the general locations of the two counterpart candidates (``C1'' and ``C2'') from \citet{cha08}, and the locations (represented by diamonds) of nearby infrared sources (Manuel A.P. Torres, private communication). }
\label{img_17091}
\end{figure}

Our \chan\ observation of \object{IGR~J17091$-$3624} coincided with the 2011 outburst so we were able to detect the source at a position of R.A. (J2000) $=$ $17^{\mathrm{h}}$ $09^{\mathrm{m}}$ $07\overset{\mathrm{s}}{.}59$ and Dec. $=$ $-36^{\circ}$ $24^{\prime}$ $25\overset{\prime\prime}{.}4$ with an error radius of $0\overset{\prime\prime}{.}6$ (90\% confidence). This is the most accurate X-ray position known for this object and it is $0\overset{\prime\prime}{.}5$ away from, and is consistent with, the XRT position which has an uncertainty of $3\overset{\prime\prime}{.}5$ \citep{ken07}. It is also compatible with the optical/IR and radio counterparts proposed by \citet{tor11} and \citet{rod11} which are both less than $0\overset{\prime\prime}{.}4$ away (their error radii are $0\overset{\prime\prime}{.}2$ at 90\% confidence, see Fig.\,\ref{img_17091}). 

It is interesting to note (as illustrated in Fig.\,\ref{img_17091}) that the IR and radio positions are only marginally compatible with each other at the 90\% confidence limit. Nevertheless, they are both inside the HRC error circle so the coordinates from the three energy bands (radio, IR, and X-rays) can be considered to be consistent. Another point worth mentioning is that the HRC error circle contains another faint star (Fig.\,\ref{img_17091}). This star, which is not one of the two blended objects that make up ``C2,'' can also be seen in the finding chart of \citet{tor11} when \object{IGR~J17091$-$3624} was in quiescence. The lack of IR variability during the X-ray outburst, and its incompatibility with either the IR or radio positions, implies that this faint star is not the counterpart to \object{IGR~J17091$-$3624}. 

Adopting an absorbed power law model with spectral parameters fixed to those of the XRT observation \citep[\nh\ $= 7.8\times10^{21}$\,\cmsq, and $\Gamma = 1.6$,][]{ken07}, the HRC count rate of 37\,\countss\ (0.3--10\,keV) converts to an absorbed flux of $2.1\times10^{-9}$\,\ergcms\ (2--10\,keV). This is a factor 35 times the 2--10-keV flux recorded by XRT \citep[$6\times10^{-11}$\,\ergcms,][]{ken07}, and nearly two orders of magnitude larger than the average flux measured by ISGRI when translated to the 2--10-keV band \citep[$3.5\times10^{-11}$\,\ergcms,][]{bir10}. The peak intensity measured by \swift-BAT\footnote{\texttt{http://heasarc.nasa.gov/docs/swift/results/transients}} on 2011 Feb. 15 (MJD\,55607) is $2.5\times10^{-9}$\,\ergcms\ (2--10\,keV). The HRC count rate converts to an observed luminosity of $1.0\times10^{38} \left [ \frac{d}{20\textrm{ kpc}} \right ]^{2}$\,\ergs.

%%__________________________________________________________________17091 Image
%\begin{figure}[!t] 
%\centering
%\includegraphics[width=0.45\textwidth,angle=0]{fig4.eps}
%\caption{HRC image of IGR~J17091$-$3624 based on an events file from standard processing which includes all \texttt{AMP\_SF} events with values between 1 and 3 (left panel), and from an events file where ``\texttt{AMP\_SF} $=$ 3'' events have been excluded (right panel). Coordinates are given as equatorial R.A. and Dec. (J2000) where North is up and East is left. A jet-like feature is visible as an extension to the NE direction in the left panel. Exclusion of events stamped as ``\texttt{AMP\_SF} $=$ 3'' removes this feature, suggesting that it is an instrumental artifact from electronic ringing \citep[see, e.g.,][]{jud00,mur00}. }
%\label{img_17091_ampsf}
%\end{figure}

\section{Undetected sources}
\label{sec_undet}

\subsection{IGR~J14043$-$6148}

We did not detect \object{IGR~J14043$-$6148} during our observation, nor were any other sources detected inside the ISGRI error circle of $4\overset{\prime}{.}5$ radius. The 3$\sigma$ upper-limit on the source count rate is 0.008\,counts\,s$^{-1}$ in the 0.3--10\,keV band. A power law whose parameters are fixed to those derived from the XRT observation \citep[\nh\ $= 7\times10^{22}$\,\cmsq, and $\Gamma = 1.8$,][]{lan11} yields an absorbed 2--10-keV flux $\leq 1.7\times10^{-12}$\,\ergcms\ which is less than the flux measured with XRT \citep[$2.9\times10^{-12}$\,\ergcms,][]{lan11} or with ISGRI \citep[$4.6\times10^{-12}$\,\ergcms,][]{bir10} when translated to the same energy range (2--10\,keV). This suggests that the source is variable in the X-rays which rules out a SNR, and points instead towards the AGN scenario proposed originally by \citet{bir10} and by \citet{lan11}.

\subsection{IGR~J16358$-$4726}

Unfortunately, \object{IGR~J16358$-$4726} was not active (or it was very faint) during our 1-ks observation and so it was not detected. No other sources were detected in the field. The non-detection is not surprising given that monitoring observations with \rxte\ \footnote{\texttt{http://asd.gsfc.nasa.gov/Craig.Markwardt/galscan}} show few periods of activity in the last 7--8 years. We set a 3$\sigma$ upper limit of 0.005\,counts\,s$^{-1}$ on the 0.3--10-keV flux from \object{IGR~J16358$-$4726}. An absorbed power law with parameters fixed to those from an \xmm\ observation in which the source was detected \citep[\nh\ $= 2\times10^{23}$\,\cmsq\ and $\Gamma = 1.5$,][]{mer06} gives an observed flux $\leq 1.8\times10^{-12}$\,\ergcms\ in the 2--10-keV band. This upper limit from \chan\ is less than the average flux observed with ISGRI when translated to the 2--10-keV band \citep[$3.9\times10^{-12}$\,\ergcms:][]{bir10}. For comparison, \citet{mer06} used \xmm\ to measure a flux of $(3.1\pm0.6)\times10^{-13}$\,\ergcms\ during detections, with 3$\sigma$ upper limits of $4\times10^{-14}$\,\ergcms\ during non-detections.

\subsection{IGR~J17597$-$2201}

\object{IGR~J17597$-$2201} was not active during our 1-ks observation, and so it was not detected. The 3$\sigma$ upper limit on the X-ray flux (0.3--10\,keV) at the \citet{rat10} source position is 0.009\,counts\,s$^{-1}$. For comparison, \chan\ recorded 0.19\,counts\,s$^{-1}$ in 2007 when the source was active \citep{rat10}. Adopting the spectral parameters from the \xmm\ observation of \citet{wal06}, i.e., \nh\ $=4.5\times10^{22}$\,\cmsq\ and $\Gamma = 1.7$, the upper limit converts to an absorbed flux (2--10\,keV) of $\leq 1.5\times10^{-12}$\,\ergcms. This is well below the average flux (2--10\,keV) extrapolated from ISGRI \citep[$3.5\times10^{-11}$\,\ergcms,][]{bir10}. The long-term light curve of \object{IGR~J17597$-$2201} from \rxte\ monitoring shows that the source was active during 2001--2008 (mostly concurrent with the ISGRI observations), and has been dormant since then. Therefore, this \chan\ upper limit represents a boundary on the quiescent flux for this source. The distance to the source is not known, but \citet{lut05} propose a distance of between 5 and 10\,kpc, while \citet{gal08} suggest an upper limit of 16\,kpc from the X-ray bursts. The \chan\ upper limit corresponds to an observed quiescent luminosity of $\le 1.8\times10^{34} \left [ \frac{d}{10\textrm{ kpc}} \right ]^{2}$\,\ergs.

\section{Summary \& Conclusions}
\label{sec_conc}

Our \chan\ observations enabled us to derive sub-arcsecond X-ray coordinates for \igrthree\ and \igrfour. The refined X-ray coordinates that obtained for \igrthree\ excludes the 2MASS star that we had previously proposed as the optical/IR counterpart (whose spectral class was the subject of disagreement), and points instead to a new (and probably distant) counterpart candidate in the mid-IR that is blended with the bright 2MASS star. For \igrfour, we provide a precise X-ray position that is consistent with the reported optical/infrared and radio counterparts, cementing its status as a microquasar. Three of our targets were not detected: \igrone, \igrtwo, and \igrfive. Nevertheless, the upper limits that we derived for their fluxes helps to establish the range of dynamic variability, which can prove useful for clarifying their nature. The non-detection of \igrone\ suggests variability that favors an active galactic nucleus rather than a supernova remnant. The upper limit for \igrfive\ sets the boundary on the quiescent flux from a probable low-mass X-ray binary that has been dormant since 2008.

\acknowledgments
The authors thank the anonymous referee whose prompt review helped improve the quality of the manuscript. Support for this work was provided by the National Aeronautics and Space Administration through \chan\ Award Number GO1-12033X issued by the \chan\ X-ray Observatory Center, which is operated by the Smithsonian Astrophysical Observatory for and on behalf of the National Aeronautics and Space Administration, under contract NAS8-03060. This research has made use of: data obtained from the High Energy Astrophysics Science Archive Research Center (HEASARC) provided by NASA's Goddard Space Flight Center; the VizieR catalogue access tool and the SIMBAD database operated at CDS, Strasbourg, France; NASA's Astrophysics Data System Bibliographic Services; the Infrared Processing and Analysis Center/California Institute of Technology, funded by the National Aeronautics and Space Administration and the National Science Foundation; the \swift/BAT transient monitor results provided by the \swift/BAT team; and the IGR Sources page (\texttt{http://irfu.cea.fr/Sap/IGR-Sources}).

{\it Facilities:} \facility{Chandra}.

\bibliographystyle{apj}
\bibliography{bodaghee.bib}
\clearpage

\end{document}